\documentclass[reprint,amsmath,aps,prb,twocolumn]{revtex4-1}
\usepackage{graphicx}
\usepackage{graphics}
\usepackage{dcolumn}
\usepackage{amsmath}
\usepackage{amssymb}
\usepackage{array}
\usepackage{multirow}
\usepackage{calc}
\usepackage{color}
\usepackage{float}
\usepackage{bm}

\begin{document}

\title{Experimental evidence for orthorhombic \textit{Fddd} crystal structure in elemental yttrium above 100 GPa}

\author{Jonathan Buhot}
\email{jonathan.buhot@bristol.ac.uk}
\author{Owen Moulding}
\author{Takaki Muramatsu}
\author{Israel Osmond}
\author{Sven Friedemann}
\email{sven.friedemann@bristol.ac.uk}

\affiliation{HH Wills Laboratory, University of Bristol, Bristol, BS8 1TL, UK}

\begin{abstract}
We present electrical resistance measurements of elemental yttrium on bulk and film samples, and both exhibit superconductivity at very high pressures. We show that the pressure dependence of the superconducting transition temperature above 100 GPa is in good agreement with the predicted $Fddd$ phase by Chen \textit{et al.} [Phys. Rev. lett. \textbf{109}, 157004 (2012)]. This result together with a new Rietveld refinement made on X-ray data at 123 GPa from Samudrala \textit{et al.} [J. Phys. Condens. Matter \textbf{24}, 362201 (2012)] offer strong evidence that the atomic structure of yttrium above 100 GPa is orthorhombic $Fddd$. Furthermore, our process of evaporating yttrium film directly on a diamond anvil is expected to be a valuable asset for future synthesis of new superhydride superconductors. 

\end{abstract}

\date{\today}

\maketitle

For almost a century, rare-earth elements have mobilized a considerable amount of effort, experimental and theoretical, for understanding their electronic structure, transport properties, and magnetic properties.
By decreasing their atomic number and under compression, most of the rare-earth elements show a common structural-transition sequence: hexagonal close-packed (hcp) $\rightarrow$ samarium-type (Sm-type) $\rightarrow$ double hexagonal close-packed (dhcp) $\rightarrow$ face-centered cubic (fcc) which is driven by an increase of the \textit{d}-band occupancy resulting from
\textit{s}-\textit{d} electron transfer \cite{duthie_correlation_1977}. 
Eventually, \textit{f} electrons get involved through hybridization which affects conduction properties and magnetic orderings and may contribute to the structure as well \cite{allen_kondo_1982, mcmahon_structure_2019}. Such studies advance our understanding of electronic correlations relevant for many materials.
In addition, hydrogen compounds of rare-earth elements lanthanum and yttrium exhibit superconductivity near room temperature under megabar pressures \cite{troyan_synthesis_2019, kong_superconductivity_2019, drozdov_superconductivity_2019-1, geballe_synthesis_2018, somayazulu_evidence_2019}. A correct description of the elemental phase of yttrium and lanthanum is important for the accurate modelling and analysis of these superconducting phases at high pressure. Whilst experimental and theoretical data agree on the high-pressure phase of lanthanum, there is no consensus yet for elemental yttrium \cite{samudrala_structural_2012, chen_predicted_2012}.

We measure resistivity and re-analyse previous X-ray data \cite{samudrala_structural_2012} to study the underlying structure and superconductivity of the high-pressure phase of yttrium.  
Elemental yttrium is known amongst the rare-earth elements as the superconductor with the highest superconducting critical temperatures ($T_{\text c}$) rising to $T_{\text{c}}=20$ K at 122 GPa \cite{hamlin_superconductivity_2007, deng_enhanced_2019}. This rise in $T_{\text c}$ is in good agreement with calculations of the distorted fcc phase up to 100 GPa. However, $T_{\text c}$ is predicted to decrease above 100 GPa after a phase transition into either a $Fddd$ or $P3_121$ structure \cite{chen_predicted_2012}. 

Yttrium undergoes a series of structural transitions under pressure: hexagonal close-packed (hcp) $\rightarrow$ samarium-type (Sm-type) $\rightarrow$ double hexagonal close-packed (dhcp) $\rightarrow$ distorted face-centered cubic (dfcc) at approximately 13, 25 and 50 GPa, respectively \cite{samudrala_structural_2012}. At about 100 GPa another structural transition has previously been observed \cite{samudrala_structural_2012} and has been refined to a $C2/m$ phase that was known, at the time, as the high pressure phase of several rare-earth metals \cite{cunningham_symmetry_2007, samudrala_high_2011, vohra_high_2008}. However, recently the $C2/m$ phase has been demonstrated to be incorrect for Nd, Tb, Gd, Dy, Ho, Er, and (probably) Tm. Instead an $oF16-Fddd$ (Pearson symbol and space group) phase has been proposed for Tb, Gd, Dy, Ho, Er, Tm and an $oF8-Fddd$ phase for Nd \cite{mcmahon_structure_2019, perreault_static_2020}. This raises questions about the conclusion of the $C2/m$ phase in yttrium \cite{samudrala_structural_2012}.
Indeed, evolutionary structure searches do not show the $C2/m$ phase among the four most energetically preferable crystal structures for yttrium at 140 GPa \cite{chen_predicted_2012}.
The $C2/m$ monoclinic structure presents strong internal stresses and even with relaxation appears to be unstable, and it converts to an orthorhombic $Cmcm$ structure \footnote{Private communication}.
Instead, \citet{chen_predicted_2012} have proposed $oF16-Fddd$ and $hP3-P3_121$ as possible stable structures for yttrium above 100 GPa and they have calculated $T_{\text c}$ associated to those phases; they predicted a suppression of $T_{\text c}$ with increasing pressure.
The $Fddd$ or $P3_121$ phases are expected to be stable up to 524 GPa where a transition to an $Fmm2$ phase has been predicted \cite{li_new_2019}. 

Here, we present resistivity measurements in bulk and film yttrium up to 164 GPa. We show that our results of $T_{\text c}$ above 100 GPa are in good agreement with the calculation for both $oF16-Fddd$ and $hP3-P3_121$ phases \cite{chen_predicted_2012}. However, we cannot discriminate between $oF16-Fddd$ and $hP3-P3_121$ phases on the basis of $T_{\text{c}}(P)$.
To do so, we have re-analysed X-ray data from \citet{samudrala_structural_2012}.
Our refinement to a $oF16-Fddd$ phase produces a good fit for the data while a $hP3-P3_121$ phase does not reproduce some key diffraction peaks.
Thus, we identify $oF16-Fddd$ as the most likely phase to consider for yttrium above 100 GPa. In addition, we compare our results on bulk and film samples.

\begin{figure}[htpb]
\centering
\includegraphics[width=1\linewidth]{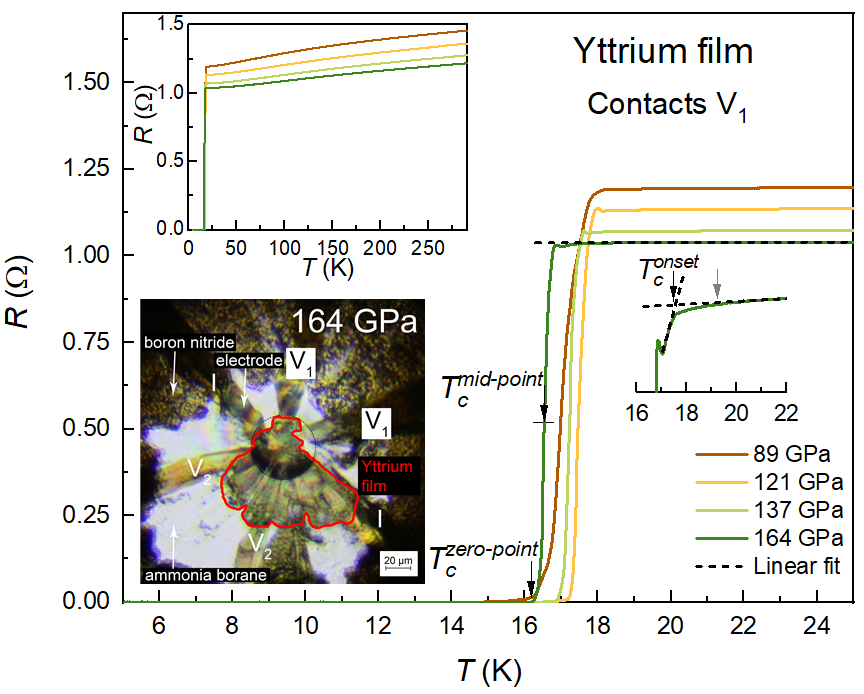}
\caption{Temperature dependence of the resistance of the yttrium film. Inset shows the full temperature range. Resistance was measured between contacts \textit{V$_1$}. Photograph shows the yttrium film highlighted in red in the pressure cell JB1 at 164 GPa. Six contacts (three pairs denoted \textit{I}, \textit{V$_1$} and \textit{V$_2$} ) are available for electrical measurements. This photograph has been taken with both transmission and reflection lighting. Initially, the sample size was about $30 \times 40$ $\mu$m$^2$. At high pressure, a part of the sample has expanded beyond the 50 $\mu$m diamond culet. The transparent region is filled with ammonia borane, and the outer brown region corresponds to the boron nitride gasket. Blacks arrows demonstrate the extraction of the superconducting critical temperatures $T_{\text{c}}^{\text{onset}}$, $T_{\text{c}}^{\text{mid-point}}$, and $T_{\text{c}}^{\text{zero-point}}$ from the resistance curve of yttrium film at 164 GPa. The grey arrow indicates where the resistance starts to deviate from the linear behaviour fitted above 22 K.
}
\label{fig1}
\end{figure}

\begin{figure}[htpb]
\centering
\includegraphics[width=1\linewidth]{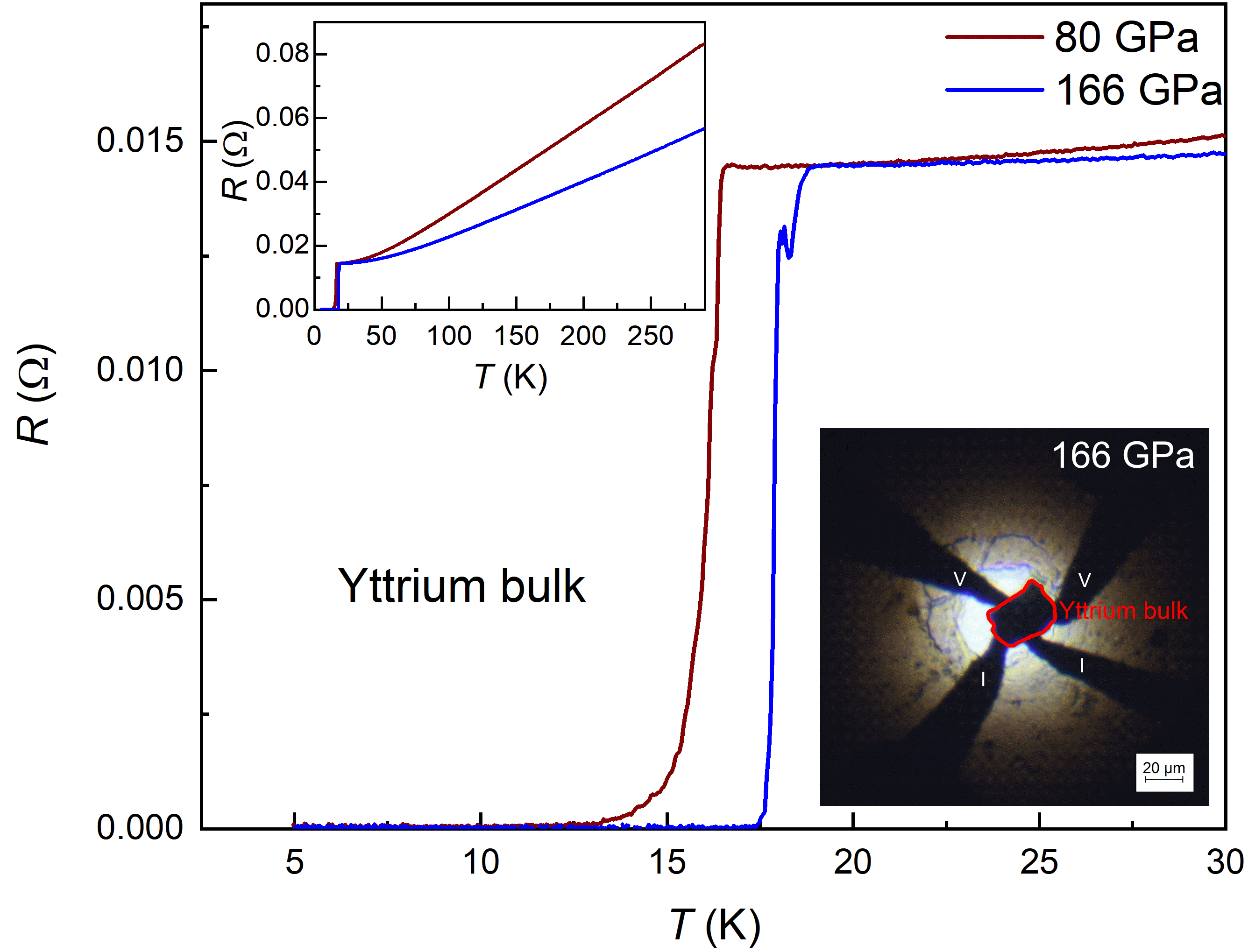}
\caption{Temperature dependence of the resistance for yttrium bulk sample TM8 of about $30 \times 20 \times 3$ $\mu $m$^3$. Inset shows the resistance curves on a full temperature range. Photograph presents the four-point configuration of the sample highlighted in red. The transparent region is filled with ammonia borane.}
\label{fig2}
\end{figure}

High pressure experiments were carried out using three home-designed diamond anvil cells (DAC) made of MP35N superalloy and equipped with 50 $\mu$m culet, type Ia and IIas (ultra low fluorescence) diamond anvils. 
Cells TM8 and TM10 were loaded with a bulk piece of yttrium of $\sim30 \times 20$ $\times 3$ $\mu$m$^3$ cut from a lump (Alfa Aesar, 99.9\% pure) and a flat yttrium wire (3 $\mu$m thickness), respectively (cf. photographs Fig. \ref{fig2} and Fig. \ref{figS2}). Cells TM8 and TM10 were prepared each with four electrodes made of flat platinum wires mounted on the anvil by hand.
Cell JB1 was loaded with an yttrium film of $\sim30 \times 40$ $\times 0.35$ $\mu$m$^3$ directly evaporated on the top of six electrodes. Here, electrodes were made by sputtering 150 nm of tungsten through a shadow mask on the anvil and then evaporating 50 nm of gold on top as a protection layer (cf. photograph Fig. \ref{fig1}); the distance between the electrode's tips was about 15 $\mu$m. 
Lumps of yttrium (Alfa Aesar, 99.9\% pure) were pre-melted in vacuum in order to release oxygen impurities. Then, the evaporation chamber was pumped twelve hours and cryo-pumped thirty minutes with a nitrogen-cooled base plate to reach a vacuum of about 1E-7 mbar.
The yttrium film was about 350 nm thick and was evaporated in approximately 35 seconds (rate of $\sim$100 $\AA$/s) at a temperature higher than the yttrium melting point.

Electrodes in cell JB1 and TM8 allowed us to measure electrical resistivity in four-point configurations while TM10 was measured using a pseudo-four-point method. Electrical resistance measurements were done with an ac-resistance bridge SIM921, Stanford Research Systems using 10 to 100 $\mu$A excitation.

High pressures between anvils were held by composite gaskets consisting of an indented piece of steel (SS 301) insulated in the center with a mixture of cubic boron nitride and epoxy.
Ammonia borane (BNH$_6$) was used in all cells as a pressure transmitting medium and also with the aim to serve as a hydrogen donor in future experiments for synthesizing new superhydride superconductors \cite{somayazulu_evidence_2019, semenok_superconductivity_2019}.
In high pressure experiments, it is well-known that hydrostaticity conditions can influence electronic and structural properties.
However, \citet{hamlin_superconductivity_2007} have shown that the superconducting transition temperatures of yttrium under non-hydrostatic pressures (without using any pressure medium) were in good agreement with data obtained in quasi-hydrostatic conditions with dense helium as a pressure medium. Even tough BNH$_6$ exhibits structural transitions at 5 and 12 GPa \cite{lin_raman_2008}, in the pressure range of our work (80 - 164 GPa), it likely offers good hydrostatic conditions. Below, we discuss the potential influence of hydrostaticity and pressure gradients in more detail. 
The pressure $p$ was determined from the shift in the high-frequency edge of the Raman spectrum of diamond, and was measured with a 532 nm solid-state laser line focused on the centre of the culet \citep{akahama_pressure_nodate}. We estimated an uncertainty of $\pm$5 GPa on the pressure determination based on the spectral resolution. Typically, we found a variation of less than 10 GPa across the culet.

We find metallic behaviour in both bulk and film samples indicative of similar quality. 
The insets of Fig. \ref{fig1} and Fig. \ref{fig2} show a decreasing $R(T)$ upon cooling. The Residual Resistivity Ratio ($\text{RRR}=\frac{R(290\text{K})}{R(20\text{K})}$) extracted for yttrium film (Cell JB1, cf. Fig. \ref{fig1}) gives a typical value of about 1.2 for all pressures, while the RRR for yttrium bulk (Cell TM8, cf. Fig. \ref{fig2}) are slightly higher with RRR $\sim$ 4 at 166 GPa, $\sim$ 8 at 80 GPa.
The difference of RRR between 166 GPa and 80 GPa for the bulk sample may come from a contribution of the Pt electrodes and a change of this resistivity contribution under pressure. Indeed, two more cells (including TM10) prepared with a bulk sample showed a RRR = 1.8 (cf Fig. \ref{figS2}) close to the value of our film sample. In summary, the RRR shows good quality of our film sample, comparable to bulk.
The SC transition is observed up to the highest measured pressure of 164 GPa and sharpens as pressure increases.
We discuss the pressure dependence of $T_{\text{c}}$ below.

Both the film and the bulk samples show superconductivity at low temperatures. Figs \ref{fig1} and \ref{fig2} show the drop to zero resistance associated with superconductivity.
Despite the spread of our film sample beyond the culet we probe superconductivity on the culet by choice of the appropriate contacts.  
$R(T)$ in the cell JB1 was measured between contacts V$_1$ connected to the film centred on the culet (cf. picture Fig. \ref{fig1}), where the pressure is maximum. For these contacts, the SC transition is sharp in temperature. 
However, a part of the sample has spread beyond the culet above 80 GPa.
$R(T)$ in this part of the sample has been measured with contacts V$_2$ (cf. picture Fig. \ref{figS1}). As expected, due to the pressure gradient beyond the culet, a very broad SC transition is observed (cf Fig. \ref{figS1}).
We analyse the SC transition measured between contacts $V_1$ and compare bulk and film in detail below.

\begin{figure}[htpb]
\centering
\includegraphics[width=1\linewidth]{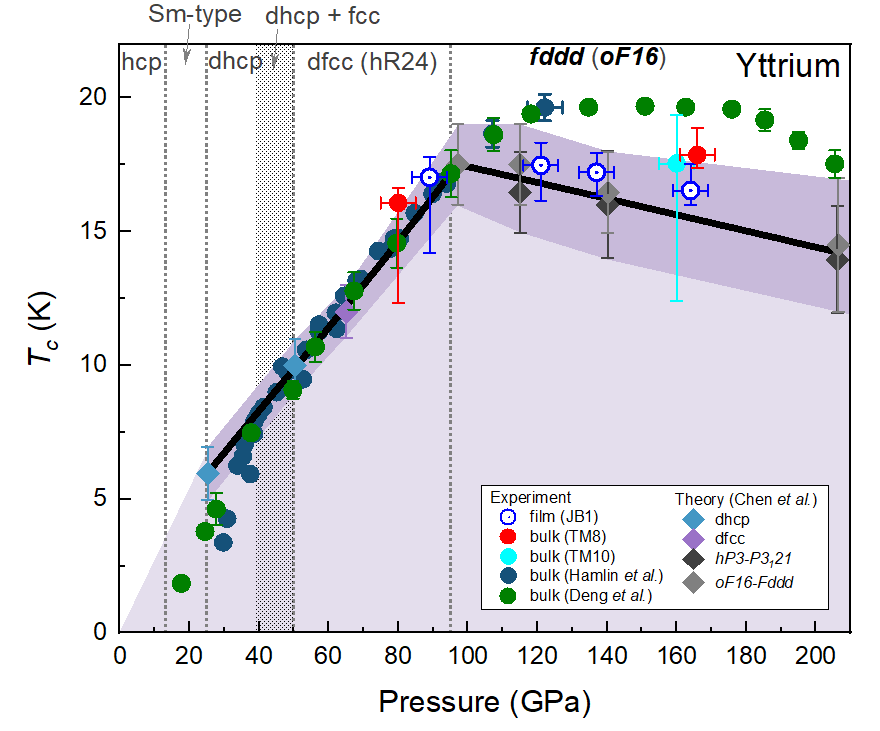}
\caption{Superconducting critical temperatures $T_{\text{c}}$ of yttrium extracted from our resistivity measurements, from the theoretical predictions of \citet{chen_predicted_2012}, from resistivity measurements of \citet{deng_enhanced_2019}, and from the ac susceptibility data (mid-point) of \citet{hamlin_superconductivity_2006, hamlin_superconductivity_2007} versus pressure.
The black line is a guide for the eye to highlight the calculated $T_\text{c}$ of \citet{chen_predicted_2012}. In refs \cite{hamlin_superconductivity_2006, hamlin_superconductivity_2007}, Hamlin \textit{et al.} measured pressure by using ruby spheres and used the ruby calibration from \cite{mao_calibration_1986, hemley_x-ray_1989}. Here, we plot the data of \citet{hamlin_superconductivity_2007} using the more recent standard ruby calibration from \cite{dewaele_compression_2008}. Vertical dashed lines correspond to the structural transitions measured by \citet{samudrala_structural_2012}.
Vertical bars on $T_{\text{c}}$ of our samples reflect the onset and zero-point of the SC transition.
}
\label{fig3}
\end{figure}

In order to closely compare yttrium bulk and film properties, we   
have extracted the onset, $T_{\text{c}}^{\text{onset}}$, the mid-point, $T_{\text{c}}^{\text{mid-point}}$, and the point where $R(T)$ reaches zero $T_{\text{c}}^{\text{zero-point}}$ as demonstrated in Fig. \ref{fig1}.
We define $T_{\text{c}}^{\text{onset}}$ as the temperature 
at which the linear behaviour above $T_{\text{c}}$ intersects with the slope where $R(T)$ drops.
$T_{\text{c}}^{\text{mid-point}}$ is defined as the temperature where the resistance is equal to $50\%$ of the normal state value, 
and $T_{\text{c}}^{\text{zero-point}}$ as the temperature for which $R=0$ within our experimental resolution of typically $0.05\%$ of the normal state resistivity.
We plot these transition temperatures together with literature data \cite{hamlin_superconductivity_2006, hamlin_superconductivity_2007, deng_enhanced_2019} in Fig. \ref{fig3}.

We observe very good agreement of $T_{\text{c}}(P)$ with previous data in the pressure range up to 100 GPa \cite{hamlin_superconductivity_2006, hamlin_superconductivity_2007, deng_enhanced_2019}.
Both our bulk and film samples follow the rising $T_{\text{c}}(P)$ in the dfcc phase below 100 GPa observed in previous studies by \citet{hamlin_superconductivity_2007}, and \citet{deng_enhanced_2019}. Furthermore, very good agreement between experimental and theoretical $T_{\text{c}}$ is found in the pressure range below 100 GPa, i.e. in the dhcp and dfcc phases \cite{chen_predicted_2012}.
  
Above 100 GPa, we observe lower values of $T_{\text{c}}$ compared to \citet{deng_enhanced_2019} for both our film and bulk samples. Whilst \citet{deng_enhanced_2019} observe a broad maximum centred around 150 GPa, we find a slight decrease of $T_{\text{c}}$ with pressure in our film samples above 100 GPa with $T_{\text{c}}$ approximately 3 K lower than \citet{deng_enhanced_2019} at 160 GPa. Our bulk samples exhibit $T_{\text{c}}$ approximately 2 K lower than \citet{deng_enhanced_2019} at 160 GPa.

Whilst pressure gradients are certainly present in our samples (and to a lesser extent in the measurements of \citet{deng_enhanced_2019} due to the larger culets used) these pressure gradients cannot easily explain the decrease in $T_{\text{c}}$ observed in our film samples above 100 GPa.
The pressure gradient in our film samples at 160 GPa is very unlikely to be large enough such that parts of the sample are at 80 GPa where $T_{\text{c}}$, according to \citet{deng_enhanced_2019}, matches our value at 160 GPa.
In fact, the pressure gradient as measured by diamond Raman across the whole sample including the parts outside the culet is less than 13 GPa.
In addition, the superior resolution of the electrical measurements on our film samples gives us high confidence that we are probing the bulk of our sample at the pressure on the culet. We find that the deviation from the normal state is less than $10^{-4}$ above $T_{\text{c}}^{\text{onset}}$ (cf. grey arrow in Fig. \ref{fig1}).
Thus we conclude that the $T_{\text{c}}$ values included in Fig. \ref{fig3} represent more than 99.99\% of the volume of our sample.
Thus, the reduction of $T_{\text{c}}$ in our film samples appears to be intrinsic. Rather the fact that the superconducting transition sharpens in both our samples and in the measurements of \citet{deng_enhanced_2019} above 100 GPa is consistent with a reduced $\dfrac{d T_{\text{c}}}{d p}$ above 100 GPa.

We cannot fully rule out that the lower $T_{\text{c}}$ in our film and bulk samples compared to data by \citet{deng_enhanced_2019} is due to impurities up to 0.1\%.
For our bulk and film samples, we have used yttrium of the same nominal purity (99.9\%, Alfa Aesar) but different origin compared to \citet{deng_enhanced_2019} (99.9\%, Ames Laboratory). In fact, \citet{deng_enhanced_2019} have shown that approximately 0.1\% magnetic impurities can reduce $T_{\text{c}}$ by 2-3 K at 150 GPa. The good agreement of $T_{\text{c}}$ of our samples (both bulk and film) with those of \citet{hamlin_superconductivity_2007} and \citet{deng_enhanced_2019} below 100 GPa indicate that our samples have impurity levels no more than 0.1\%.
This level of 0.1\% is just within the limits of the purity of our yttrium if all impurities were magnetic. We note that we can rule out additional contamination with magnetic impurities to our film samples during evaporation as no magnetic materials have been evaporated with this vacuum system before.
  
In summary, we observe a slightly lower $T_{\text{c}}$ compared to \citet{deng_enhanced_2019}. Yet, both our measurements and those of \citet{deng_enhanced_2019} show a marked deviation from the linear increase of $T_{\text{c}}(P)$ observed in the dfcc phase between 50 GPa to 100 GPa.  In fact, our film samples show a reduction of $T_{\text{c}}(P)$ above 100 GPa in excellent agreement with the values calculated for the predicted $Fddd$ and $P3_121$ phases \cite{chen_predicted_2012}. Whilst this agreement might be coincidental, the general trend of $T_{\text{c}}(P)$ observed by us and \citet{deng_enhanced_2019} is in rough agreement with the calculation. In order to distinguish between the two candidate phases, we re-analyse previous XRD data by \citet{samudrala_structural_2012} below.

\begin{figure}[htpb]
\centering
\includegraphics[width=1\linewidth]{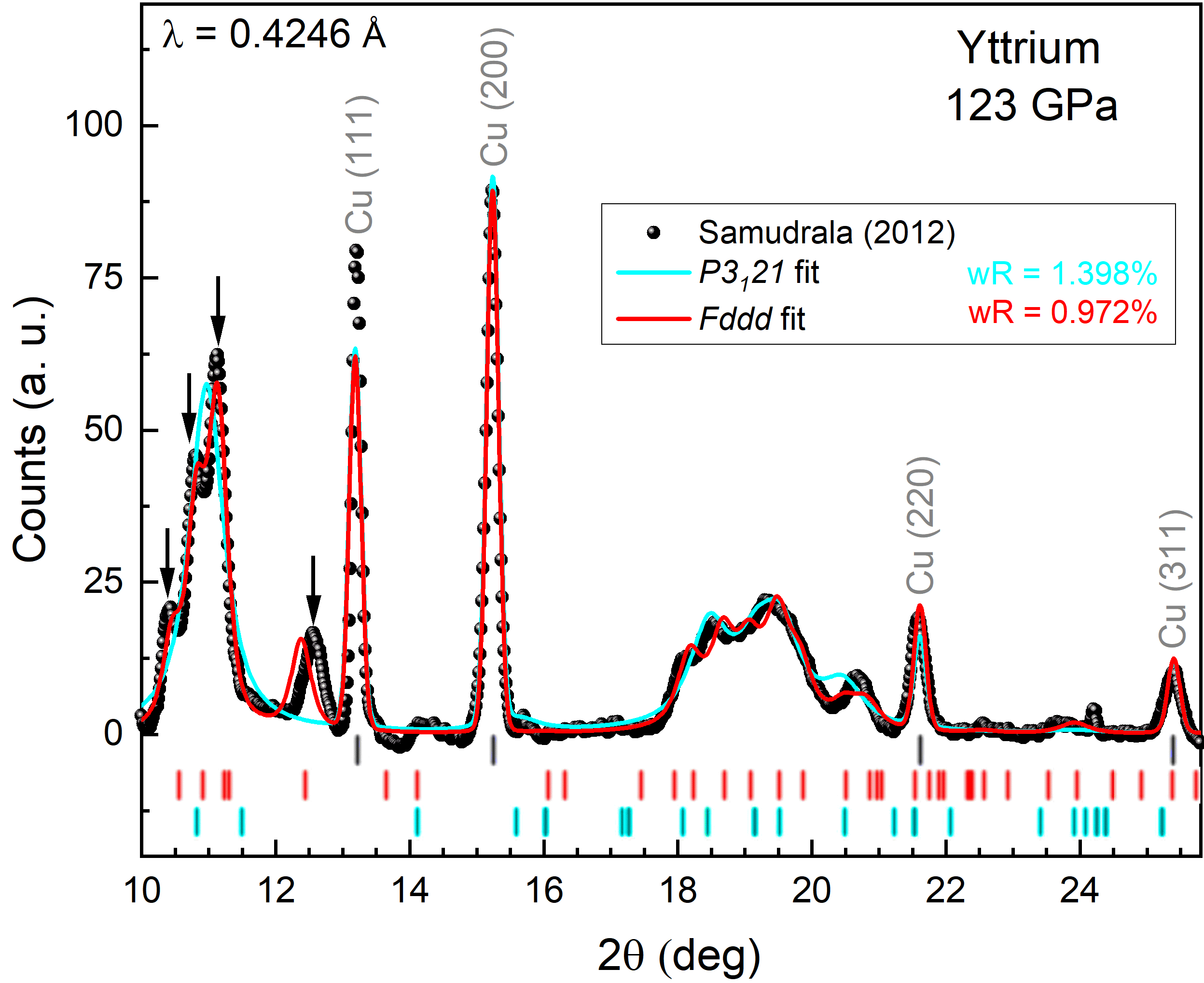}
\caption{Rietveld refinements of yttrium at 123 GPa to both $P3_121$, and $Fddd$ structures.
Experimental X-ray diffraction pattern (black spheres) are from \citet{samudrala_structural_2012}. The copper pressure marker peaks are indexed. The grey, red and light blue ticks indicate the diffraction peaks corresponding to the copper, $Fddd$ and $P3_121$ phases, respectively. The structural refinement parameters are included in the text.
}
\label{fig4}
\end{figure}

We have shown that our experimental values of $T_{\text c}$ match well the predicted ones for $oF16-Fddd$ and $hP3-P3_121$. However, the similar $T_{\text c}$ predictions for $oF16-Fddd$ and $hP3-P3_121$ prevent us from discriminating between the two phases.
To do so, we have re-analysed the X-ray data of \citet{samudrala_structural_2012} with both $oF16-Fddd$ and $hP3-P3_121$ structures as shown in Fig. \ref{fig4}.
The experimental X-ray diffraction pattern clearly shows the presence of at least four peaks between $2\theta=10^{\circ}$ and $2\theta=13^{\circ}$ which can be refined with the $oF16-Fddd$ phase but not with the $hP3-P3_121$ which produce only two peaks over this range.
Therefore, we can rule out $hP3-P3_121$ as the structural phase above 100 GPa. Moreover, the weighted residual (wR) coefficient, resulting from the fit, is lower for $oF16-Fddd$ than $hP3-P3_121$, indicating a better fit with $oF16-Fddd$. 
The lattice parameters obtained are \textit{a} = 4.475(2) \AA, \textit{b} = 2.762(1) \AA, \textit{c} = 18.023(6) \AA, $\alpha=\beta=\gamma=90^{\circ}$ and $V=13.92(1)$ \AA$^3$/atom for the $oF16-Fddd$ phase. It is worth noting that the volume per atom obtained with this $oF16-Fddd$ refinement is comparable with the volume $V=13.73$ \AA$^3$/atom found in ref. \cite{samudrala_structural_2012} with a $C2/m$ refinement. 
At 102 GPa (cf. Fig \ref{figS3}), our re-analysis of the X-ray data gives lattice parameters \textit{a} = 4.859(4) \AA, \textit{b} = 2.791(3) \AA, \textit{c} = 18.02(2) \AA, $\alpha=\beta=\gamma=90^{\circ}$ and $V=15.27(1)$ \AA$^3$/atom in very good agreement with the predicted lattice parameters \textit{a} = 4.81 \AA, \textit{b} = 2.83 \AA, \textit{c} = 17.70 \AA, $\alpha=\beta=\gamma=90^{\circ}$ and $V=15.06$ \AA$^3$/atom at 97 GPa of \citet{chen_predicted_2012}. Consequently, we now interpret the X-ray data results as evidence for the $oF16-Fddd$ phase at pressures above 100 GPa.
Thus, yttrium shows the same high pressure phase like Tb, Gd, Dy, Ho, Er, Tm. This suggests similar electronic orbitals determining the crystal structure for yttrium and lanthanide elements.
The sequence of structural transitions in rare-earth elements is linked to the \textit{s}-\textit{d} transfer of electronic states \cite{duthie_correlation_1977}. Yet, a partial hybridisation of \textit{4f} electrons has been identified in Gd and Tb at high pressures with a partial electron transfer \cite{mcmahon_structure_2019}. For Gd, Dy, Sm, and Nd, magnetic ground states result from remaining local \textit{f} states with the help of the Ruderman-Kittel-Kasuya-Yosida-interaction mediated by conduction electrons \cite{mcmahon_structure_2019, deng_enhanced_2019}. For Tb, hybridisation of delocalised \textit{f} moments with conductions electrons has been suggested to stabilise the $oF16-Fddd$ structure at high pressures \cite{mcmahon_structure_2019}. In yttrium, the \textit{f} shells are empty and cannot be occupied in the pressure range studied here. Thus, high-pressure structure studies of yttrium offer to study the bare \textit{s}-\textit{d} transfer. Indeed, we can rule out magnetic order and fluctuating magnetic moments in the accessed pressure range as this would strongly suppress superconductivity. Thus, our results indicate that the $oF16-Fddd$ phase can also be stabilised without partially occupied \textit{f} states.


In summary, we show that both superconducting temperatures in yttrium under megabar pressures and re-analysis of X-ray results provide experimental evidence of the predicted $Fddd$ phase. This result also highlights the reliability of structure search methods and calculations of superconductivity (using evolutionary algorithms in conjunction with density functional theory) and encourage such advanced theoretical work in searching for novel high-pressure phases. Furthermore, the electronic properties of yttrium film is found to be similar to yttrium bulk sample. This opens a new route in the quest for studying new room temperature superconductors, such as such as YH$_9$, amongst other superhydrides. Indeed, film evaporation would allow a more accurate control of the sample size and would help to reach the ideal stoichiometry in the synthesis of new superhydride superconductors. 

\begin{acknowledgments}

The authors thank Gopi K Samudrala and Yogesh K Vohra for sharing X-ray data from their publication \cite{samudrala_structural_2012} and for valuable discussion. The authors also thank Chris Pickard and James Schilling for very fruitful discussions. The authors acknowledge support by the EPSRC under grants EP/L015544/1 and funding from the European Research Council (ERC) under the European Union's Horizon 2020 research and innovation programme Grant agreement No.715262-HPSuper).

\end{acknowledgments}

\section*{APPENDIX: MATERIALS, METHODS, AND ADDITIONAL SUPPORTING INFORMATION}

\setcounter{figure}{0}
\renewcommand{\thefigure}{S\arabic{figure}}

This appendix provides additional information on pressure cells as well as further data.

\begin{figure}[htpb]
\centering
\includegraphics[width=1\linewidth]{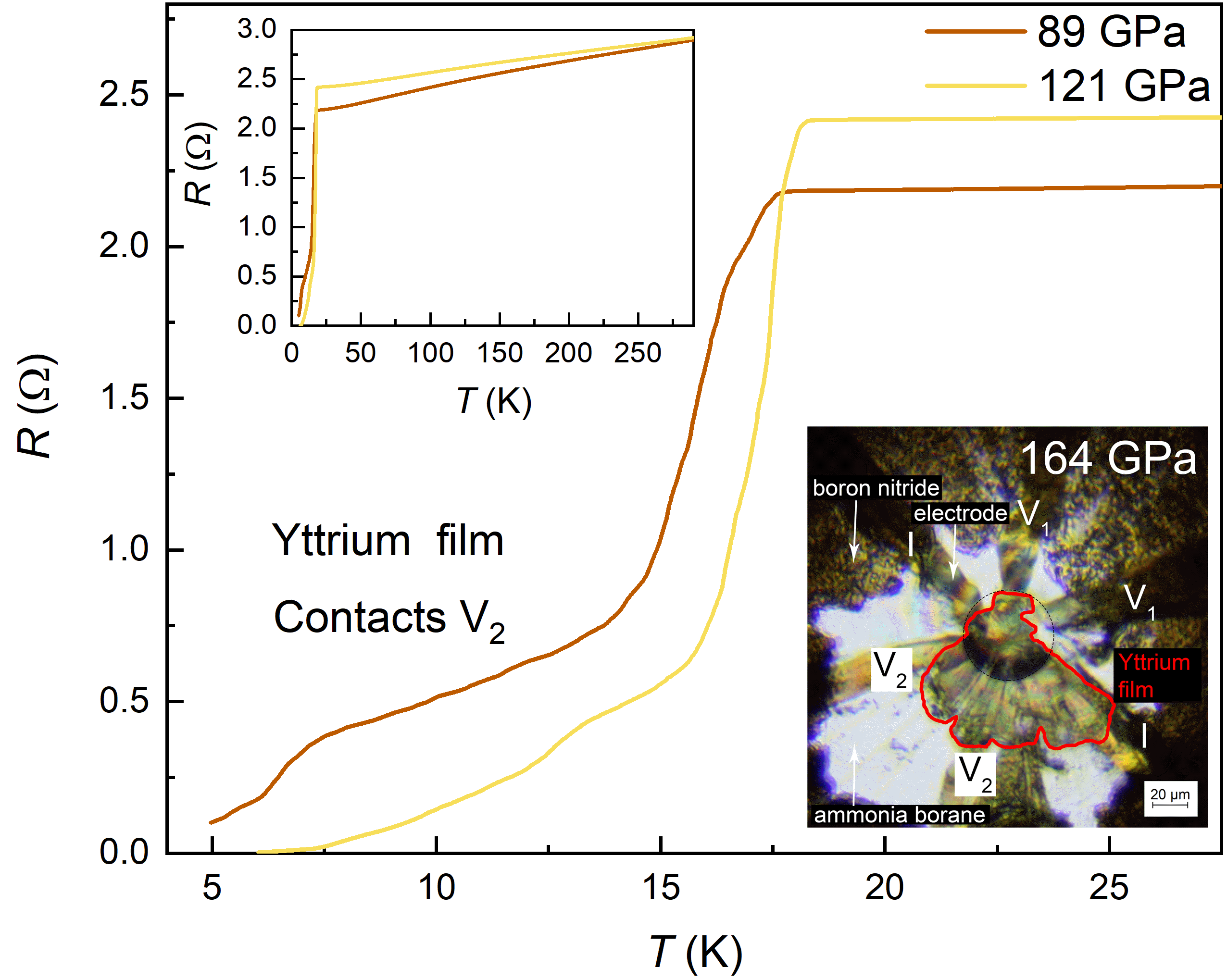}
\caption{Temperature dependences of the resistance on contacts V$_2$ of yttrium film in the pressure cell JB1. Inset shows the resistance curves on a full temperature range. Photograph presents the configuration of the measurements.}
\label{figS1}
\end{figure}

\begin{figure}[htpb]
\centering
\includegraphics[width=1\linewidth]{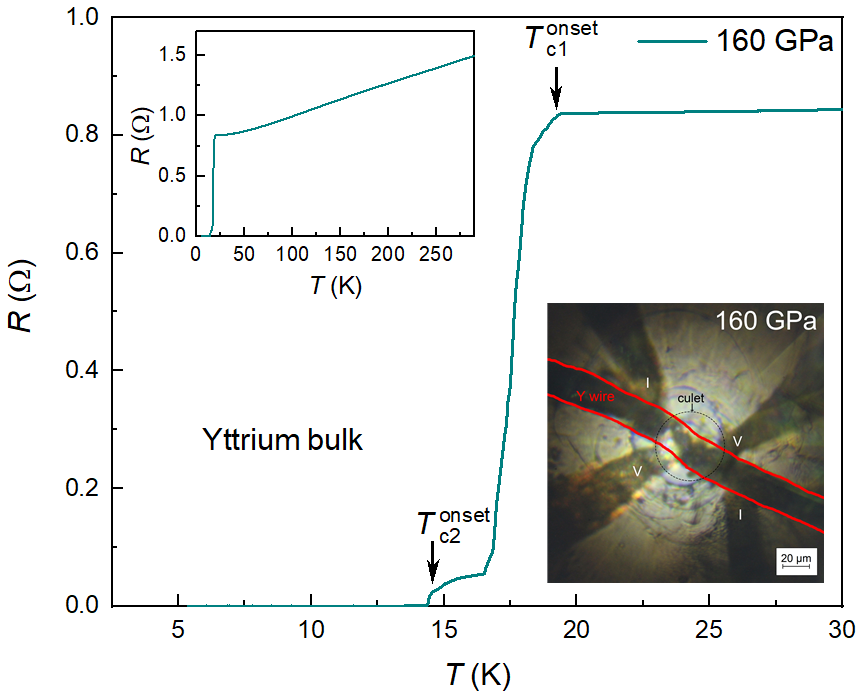}
\caption{Temperature dependence of the resistance for a flat yttrium wire of about 3 $\mu$m  thickness in the pressure cell TM10 at 160 GPa. Inset shows the resistance curve on a full temperature range. Photograph presents the pseudo four-points configuration of the measurements. Arrows indicate the two superconducting transitions due to a gradient of pressure between the voltage electrodes.}
\label{figS2}
\end{figure}

\begin{figure}[htpb]
\centering
\includegraphics[width=1\linewidth]{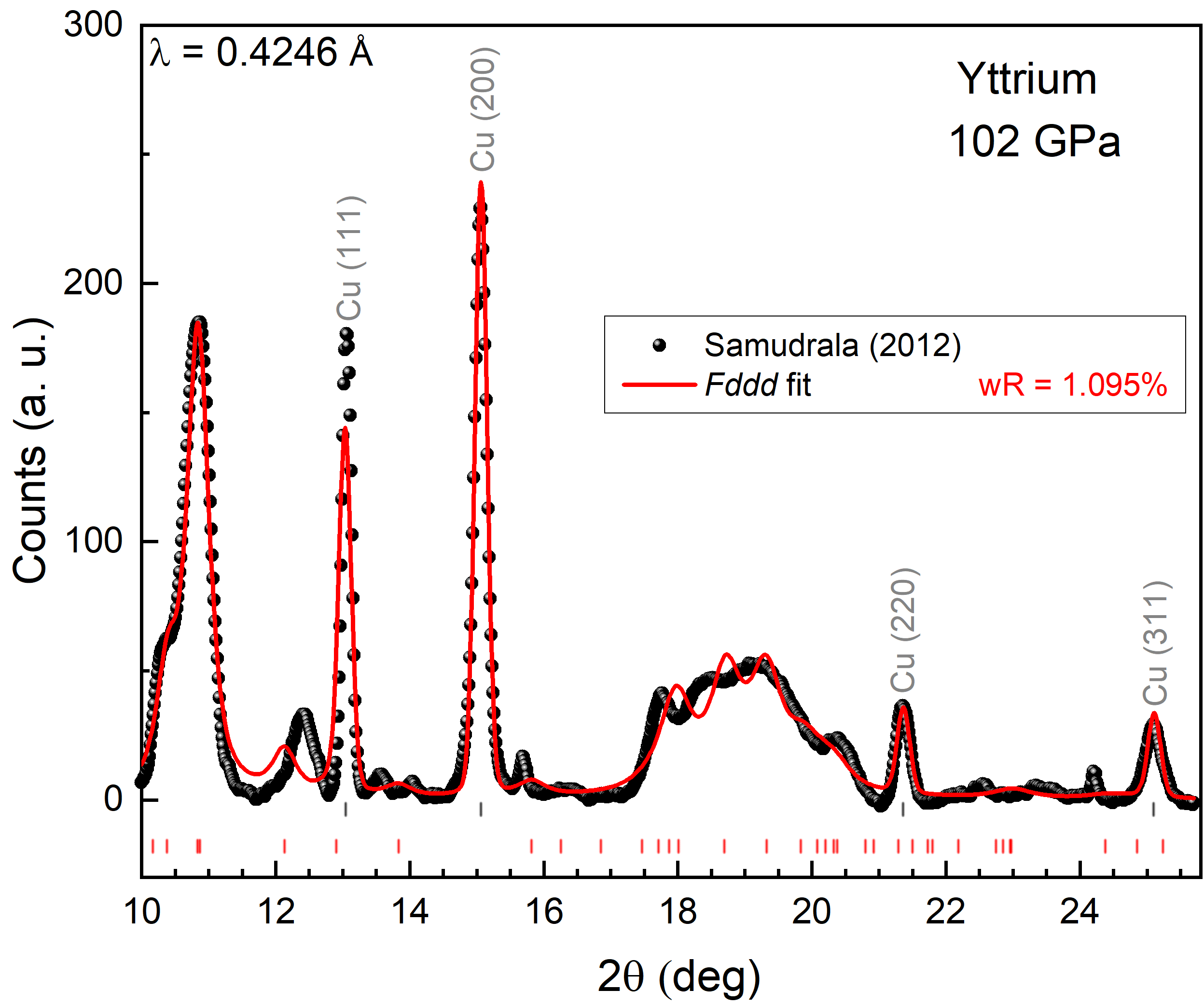}
\caption{Rietveld refinement of yttrium at 102 GPa to a Fddd structure.
Experimental X-ray diffraction pattern (black spheres) comes  from \citet{samudrala_structural_2012}. The copper pressure marker peaks are indexed. The grey and red ticks indicate the diffraction peaks corresponding to the copper and $Fddd$ phase, respectively. The structural refinement parameters are included in the text.}
\label{figS3}
\end{figure}

\clearpage

\bibliography{biblio}
\bibliographystyle{apsrev4-2}
\end{document}